\documentclass{article}
\usepackage{amsmath, amssymb, graphicx}
\usepackage{booktabs}
\usepackage{amsthm} 
\usepackage{hyperref}
\usepackage{algorithm}
\usepackage{algpseudocode}
\usepackage[numbers]{natbib}
\usepackage{booktabs} 
\usepackage{array}    
\usepackage{caption}
\usepackage[a4paper, total={6in, 8in}, left=0.5in, right=0.5in, top=1in, bottom=1in]{geometry} 
\usepackage{algorithm}
\usepackage{algorithmicx}
\usepackage{algpseudocode}
\usepackage{color}
\usepackage{tikz}
\usetikzlibrary{positioning, arrows.meta, shapes.geometric}
\usepackage{multirow}

\usepackage[below]{placeins}
\usepackage{sidecap}
\usepackage{graphicx}
\usepackage{moreverb}   
\usepackage{hyperref}
\usepackage{fancyhdr}
\usepackage[latin1]{inputenc}
\usepackage{wrapfig}
\usepackage[numbers]{natbib}

\numberwithin{equation}{section}

\title{Deep Generative Models for Synthetic Financial Data: Applications to Portfolio and Risk Modeling}
\author{Christophe D. Hounwanou \thanks{christophe.hounwanou@aims.ac.rw} \\ 
        African Institute for Mathematical Sciences, AIMS Rwanda \\ 
         \\
        \and
        Ya\'e Ulrich Gaba\thanks{yaeulrich.gaba@gmail.com} \\ 
        Sefako Makgatho Health Sciences University (SMU)\\ Pretoria, South Africa \\
        \&\\
         AI Research and Innovation Nexus for Africa (AIRINA Labs) \\ AI.Technipreneurs, B\'enin
}

\begin{document}

\maketitle
\begin{abstract}
Synthetic financial data provides a practical solution to the privacy, accessibility, and reproducibility challenges that often constrain empirical research in quantitative finance. This paper investigates the use of deep generative models, specifically Time-series Generative Adversarial Networks (TimeGAN) and Variational Autoencoders (VAEs) to generate realistic synthetic financial return series for portfolio construction and risk modeling applications. Using historical daily returns from the S\&P 500 as a benchmark, we generate synthetic datasets under comparable market conditions and evaluate them using statistical similarity metrics, temporal structure tests, and downstream financial tasks. The study shows that TimeGAN produces synthetic data with distributional shapes, volatility patterns, and autocorrelation behaviour that are close to those observed in real returns. When applied to mean--variance portfolio optimization, the resulting synthetic datasets lead to portfolio weights, Sharpe ratios, and risk levels that remain close to those obtained from real data. The VAE provides more stable training but tends to smooth extreme market movements, which affects risk estimation. Finally, the analysis supports the use of synthetic datasets as substitutes for real financial data in portfolio analysis and risk simulation, particularly when models are able to capture temporal dynamics. Synthetic data therefore provides a privacy-preserving, cost-effective, and reproducible tool for financial experimentation and model development.
\end{abstract}

\section{Introduction}

The widespread adoption of data-driven methods in finance has been supported by recent progress in machine learning and deep learning for portfolio optimization, trading, and risk management \cite{gu2020empirical, heaton2017deep}. However, financial datasets are often scarce, proprietary, and sensitive, which restricts reproducibility and limits the development of scalable research frameworks \cite{li2023evaluation}. Furthermore, the stochastic and non-stationary nature of financial markets adds complexity to model training and evaluation \cite{tsay2010analysis}.

\vspace{0.3cm}
\noindent
Synthetic data generation has emerged as a promising approach to address these challenges. Advances in generative modeling, including Generative Adversarial Networks (GANs) and Variational Autoencoders (VAEs), allow the production of realistic and privacy-preserving data that retain the key statistical properties of financial markets \cite{goodfellow2014generative, kingma2014auto}. In particular, models such as TimeGAN \cite{yoon2019time}, which combine recurrent architectures with adversarial learning, show strong potential for generating temporal sequences with coherent market dynamics.

\vspace{0.3cm}
\noindent
Despite increasing interest, there is limited evidence on the practical utility of synthetic financial time series in downstream applications such as portfolio construction, risk estimation, and strategy backtesting \cite{li2023evaluation, takahashi2019data}. In other words, it remains unclear whether synthetic data can reliably replicate market behavior and support quantitative decision-making tasks.

\vspace{0.3cm}
\noindent
\textbf{Research Question:} \textit{This study investigates whether synthetic financial time series generated by TimeGAN and VAE can faithfully reproduce the statistical and temporal properties of real market data, and whether these synthetic datasets can support downstream tasks such as portfolio optimization, risk estimation, and backtesting.}

\vspace{0.3cm}
\noindent
We focus on S\&P~500 data as a benchmark, generating synthetic sequences using TimeGAN and VAEs, and evaluate their ability to reproduce key market patterns. The empirical results provide insight into the fidelity and practical utility of synthetic financial data, with implications for privacy-preserving and reproducible research in quantitative finance.

\vspace{0.3cm}
\noindent
The remainder of this paper is organized as follows. Section~\ref{section2} introduces the theoretical background and fundamentals of synthetic data generation. Section~\ref{section3} presents the methodology and experimental setup used for portfolio and risk modeling. Section~\ref{section4} reports the empirical results. Section~\ref{section5} discusses the main implications and limitations. Section~\ref{section6} concludes with perspectives for future research.

\section{Preliminaries}\label{section2}

This section introduces the conceptual background required to analyze the contribution of synthetic financial data to portfolio optimization and risk modeling. While the first paper focused on the generation and statistical validation of synthetic time series, the present work examines how such data can support decision-oriented tasks. We begin by formalizing the standard portfolio and risk modeling framework, then outline how synthetic datasets can be integrated into estimation and evaluation procedures.

\subsection{Formal setting: Portfolio and risk modeling framework}

Let $\mathbf{r}_t = (r_{1,t}, r_{2,t}, \ldots, r_{N,t})^\top$ denote the vector of asset log-returns at time $t$, where $N$ is the number of assets.  
A portfolio is characterized by a weight vector $\mathbf{w} = (w_1, w_2, \ldots, w_N)^\top$ satisfying the budget constraint $\sum_{i=1}^N w_i = 1$.  
The portfolio return at time $t$ is:
\[
R_t = \mathbf{w}^\top \mathbf{r}_t,
\]
with expected return and variance given by:
\[
\mu_p = \mathbb{E}[R_t], 
\qquad 
\sigma_p^2 = \mathbb{V}[R_t] = \mathbf{w}^\top \Sigma \mathbf{w},
\]
where $\Sigma$ denotes the covariance matrix of asset returns.

\noindent
The classical mean-variance optimization problem of Markowitz (1952) can be written as:
\[
\min_{\mathbf{w}} \ \mathbf{w}^\top \Sigma \mathbf{w}
\quad \text{subject to} \quad 
\mathbf{w}^\top \mathbf{1} = 1, 
\qquad 
\mathbf{w}^\top \mu = \mu_p^*,
\]
where $\mu_p^*$ is a target expected return.  
In practice, the quantities $\mu$ and $\Sigma$ are estimated from historical data, which can lead to instability and sensitivity to sampling variation, particularly in settings with limited data or pronounced non-stationarity.

\noindent
Synthetic data generation offers a potential way to address these challenges by providing additional samples that reproduce key temporal and distributional characteristics of financial time series. Such augmented datasets can be used to improve the robustness of parameter estimation, explore alternative market scenarios, and support stress-testing procedures in portfolio and risk modeling.

\subsection{Synthetic data as a bridge between privacy and utility}

Synthetic financial data provides a practical approach to address two common challenges in quantitative finance: limited data access and the need for privacy-preserving analytics.  
Financial datasets are often restricted due to confidentiality concerns, making reproducibility and robust testing of models difficult. Synthetic data, generated via models such as Variational Autoencoders (VAEs) or Time-series GANs (TimeGAN), offers a way to simulate realistic market behavior without exposing sensitive information.

\noindent
Formally, let a real dataset be 
\(\mathcal{D} = \{\mathbf{r}_{1:T}^{(i)}\}_{i=1}^M\), 
where each \(\mathbf{r}_{1:T}^{(i)}\) is a sequence of log-returns for \(N\) assets over \(T\) time steps.  
A generative model \(G_\theta\) maps latent variables \(\mathbf{z} \sim p(\mathbf{z})\) to synthetic sequences:
\[
\tilde{\mathbf{r}}_{1:T} = G_\theta(\mathbf{z})
\]
such that the generated distribution \(p_\theta(\tilde{\mathbf{r}}_{1:T})\) approximates the real data distribution \(p(\mathbf{r}_{1:T})\) according to a divergence measure (e.g., Kullback-Leibler, Wasserstein).  

This approach allows researchers to:
\begin{itemize}
    \item Augment small datasets for more stable estimation of portfolio and risk parameters,
    \item Conduct stress-testing under alternative market scenarios,
    \item Share data and models across teams or institutions without exposing real transaction records.
\end{itemize}

\noindent
Synthetic data connects privacy and practical use, allowing classical portfolio and risk modeling to be applied in a reproducible and secure way.

\subsection{Synthetic data for downstream portfolio and risk tasks}

Beyond generating statistically realistic time series, the ultimate goal of synthetic financial data is to support practical decision-making tasks.  
These include portfolio optimization, risk assessment, and stress-testing, where reliable estimates of returns, covariances, and extreme events are essential.  \\

\noindent
Let \(\tilde{\mathbf{r}}_t\) denote synthetic asset returns at time \(t\), generated from a model trained on real data \(\mathbf{r}_t\).  
We define the following key downstream objectives:

\begin{itemize}
    \item \textbf{Portfolio optimization:} Use synthetic returns to compute expected returns \(\tilde{\mu}\) and covariance \(\tilde{\Sigma}\), which then inform mean-variance or risk-parity allocations.  
    \item \textbf{Risk modeling:} Evaluate whether synthetic series reproduce key risk measures, including variance, Value-at-Risk (VaR), and conditional VaR.  
    \item \textbf{Stress testing and scenario analysis:} Simulate rare or extreme market events in synthetic data to examine the robustness of portfolios or trading strategies.  
\end{itemize}

\noindent
Formally, the downstream task evaluation examines whether using \(\tilde{\mathbf{r}}_t\) yields decision metrics close to those obtained from the real data \(\mathbf{r}_t\), for example:

\[
\tilde{\mu}_p \approx \mu_p, \quad
\tilde{\sigma}_p^2 \approx \sigma_p^2, \quad
\text{and portfolio allocations } \tilde{\mathbf{w}} \approx \mathbf{w}.
\]

\noindent
This section sets the stage for our methodology, where synthetic datasets will be assessed not only on statistical fidelity but also on their practical utility in realistic financial tasks.

\subsection{Evaluation dimensions: fidelity, utility, and robustness}

Assessing synthetic financial data requires a multi-dimensional perspective that captures both statistical quality and practical relevance. Building on the previous subsections, we define three key evaluation dimensions:

\begin{itemize}
    \item \textbf{Fidelity:} Measures how closely the synthetic series \(\tilde{\mathbf{r}}_t\) replicates the statistical and temporal characteristics of real returns \(\mathbf{r}_t\). This includes first- and higher-order moments, autocorrelation, volatility clustering, and distributional shape. High fidelity ensures that synthetic data reflects the essential market dynamics needed for realistic modeling.

    \item \textbf{Utility:} Quantifies the effectiveness of synthetic data in supporting downstream tasks such as portfolio optimization, risk estimation, and stress-testing. Specifically, we compare performance metrics derived from synthetic series (e.g., expected return \(\tilde{\mu}_p\), variance \(\tilde{\sigma}_p^2\), Value-at-Risk, portfolio allocations \(\tilde{\mathbf{w}}\)) with those computed from real data. High utility implies that decisions based on synthetic data are consistent with those based on real data.

    \item \textbf{Robustness:} Evaluates the stability and reliability of synthetic datasets across different market regimes, random seeds, or slight perturbations in input data. A robust synthetic model maintains consistent statistical properties and downstream task performance under varying conditions, which is crucial for stress-testing, scenario analysis, and generalization to unseen market events.
\end{itemize}

\noindent
These three dimensions form a clear framework for evaluating synthetic financial data both as a statistical representation and as a practical tool for financial decision-making.  
The next section describes the methodology for generating and testing synthetic datasets in portfolio construction and risk analysis.

\section{Methodology}\label{section3}

This section details the methodological framework used to evaluate the usability of synthetic financial datasets in practical quantitative finance tasks. Our central research question is:

\textit{ Can synthetic financial time series generated by TimeGAN and VAE reliably replicate the statistical and temporal properties of real market data and support downstream tasks such as portfolio optimization, risk estimation, and backtesting? }

\noindent
The experimental pipeline consists of four stages: data acquisition and preprocessing, synthetic data generation, downstream modeling, and comparative evaluation.

\subsection{Dataset and preprocessing}

The empirical study uses daily closing prices of the S\&P 500 index from January 2000 to June 2024. The dataset contains only the index, and all downstream analyses are based on its log-returns. Raw prices are transformed into log-returns to ensure stationarity and comparability:

\[
r_t = \ln \left( \frac{P_t}{P_{t-1}} \right),
\]

where \(P_t\) is the adjusted closing price at time \(t\), and \(r_t\) is the daily log-return.

\noindent
Stationarity is verified using the Augmented Dickey-Fuller (ADF) test. The series is then standardized to zero mean and unit variance before being input into the generative models. To ensure robustness, we also examine alternative rolling-window lengths (\(T = 10, 20, 60\) days) and verify that preprocessing choices do not significantly affect downstream results.

\begin{table}[H]
\centering
\caption{Summary statistics of S\&P 500 daily log-returns (2000-2024)}
\begin{tabular}{lcccc}
\toprule
Statistic & Mean & Std. Dev. & Skewness & Kurtosis \\
\midrule
Real Data & 0.00041 & 0.0127 & -0.45 & 7.88 \\
\bottomrule
\end{tabular}
\label{table:summary_stats}
\end{table}

\noindent
This preprocessing framework ensures that the data fed to the generative models is statistically consistent, preserving key distributional and temporal properties of the original S\&P 500 returns.

\subsection{Synthetic data generation}

Synthetic datasets were produced using two state-of-the-art generative models: \textbf{TimeGAN}~\cite{yoon2019timegan} and \textbf{Variational Autoencoder (VAE)}~\cite{kingma2013auto}. TimeGAN combines recurrent neural networks with adversarial training to capture both temporal dynamics and feature correlations, while VAEs learn a probabilistic latent representation to generate smooth and continuous synthetic sequences.

\noindent
Other generative approaches, such as diffusion models, WGANs, or Copula-based models, were excluded in this study due to either high computational costs or scope limitations, allowing us to focus on two widely studied and complementary paradigms.

\subsubsection{Model training and hyperparameters}
For each model, training was performed on the preprocessed S\&P 500 log-returns using the following configurations:
\begin{itemize}
    \item TimeGAN: 200 epochs, learning rate 0.001, batch size 64.
    \item VAE: 150 epochs, learning rate 0.001, batch size 128.
    \item Random seeds were fixed to ensure reproducibility.
\end{itemize}
\noindent
Convergence was monitored through model-specific loss curves, and latent-space representations were visually inspected to verify stability and meaningful structure.

\subsubsection{Pipeline overview}
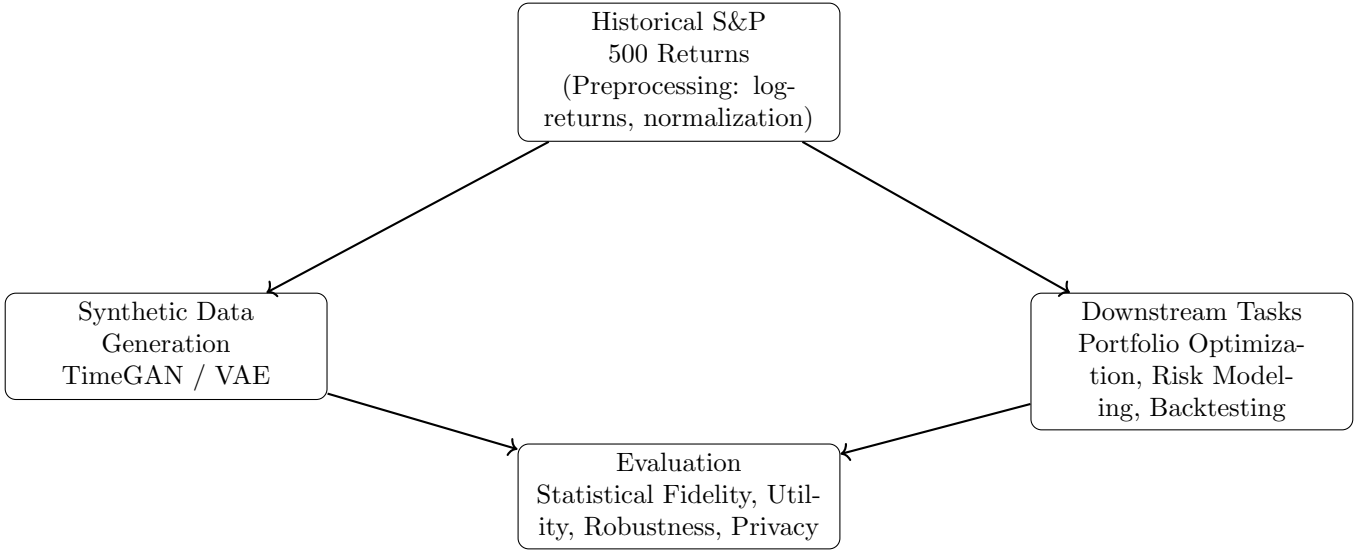
\begin{figure}[H]
\centering
\begin{tikzpicture}[node distance=2.5cm, auto]
    \node[draw, rectangle, rounded corners, text width=4cm, align=center] (data) {Historical S\&P 500 Returns \\ (Preprocessing: log-returns, normalization)};
    \node[draw, rectangle, rounded corners, text width=4cm, align=center, below left=2cm and 2.5cm of data] (gen) {Synthetic Data Generation \\ TimeGAN / VAE};
    \node[draw, rectangle, rounded corners, text width=4cm, align=center, below right=2cm and 2.5cm of data] (tasks) {Downstream Tasks \\ Portfolio Optimization, Risk Modeling, Backtesting};
    \node[draw, rectangle, rounded corners, text width=4cm, align=center, below=4cm of data] (eval) {Evaluation \\ Statistical Fidelity, Utility, Robustness, Privacy};
    \draw[->, thick] (data) -- (gen);
    \draw[->, thick] (data) -- (tasks);
    \draw[->, thick] (gen) -- (eval);
    \draw[->, thick] (tasks) -- (eval);
\end{tikzpicture}
\caption{End-to-end pipeline for synthetic financial data generation and downstream evaluation.}
\label{fig:synthetic_pipeline}
\end{figure}

Figure~\ref{fig:synthetic_pipeline} illustrates the end-to-end workflow for generating and evaluating synthetic financial datasets.

\noindent
This setup ensures that downstream tasks can be tested on data that closely mimics real market behavior without compromising confidentiality.

\subsection{Portfolio optimization framework}

To assess the usability of synthetic financial data, we adopt the classical \textit{mean-variance} optimization framework~\cite{markowitz1952portfolio}, extended here to a multi-asset synthetic dataset scenario. This allows us to evaluate whether synthetic data can reliably reproduce realistic portfolio allocations.

\subsubsection{Optimization problem}

Let \( \boldsymbol{w} = (w_1, w_2, \dots, w_N)^\top \) denote portfolio weights, \( \boldsymbol{\mu} \) the expected return vector, and \( \Sigma \) the covariance matrix. The optimization problem is formulated as:

\[
\boxed{
\min_{\boldsymbol{w}} \; \boldsymbol{w}^\top \Sigma \boldsymbol{w} 
\quad \text{s.t.} \quad 
\boldsymbol{w}^\top \boldsymbol{\mu} = \mu_p, \quad 
\boldsymbol{w}^\top \mathbf{1} = 1, \quad w_i \ge 0
}
\]

The Lagrangian formulation introduces multipliers \(\lambda, \gamma\) for the constraints:

\[
\mathcal{L}(\boldsymbol{w}, \lambda, \gamma) = \boldsymbol{w}^\top \Sigma \boldsymbol{w} 
- \lambda (\boldsymbol{w}^\top \boldsymbol{\mu} - \mu_p) 
- \gamma (\boldsymbol{w}^\top \mathbf{1} - 1)
\]

Solving the first-order conditions yields the optimal weights:

\[
\boldsymbol{w}^* = \Sigma^{-1} \left( \frac{\lambda}{2} \boldsymbol{\mu} + \frac{\gamma}{2} \mathbf{1} \right)
\]

\[
\begin{bmatrix}
\boldsymbol{\mu}^\top \Sigma^{-1} \boldsymbol{\mu} & \boldsymbol{\mu}^\top \Sigma^{-1} \mathbf{1} \\
\mathbf{1}^\top \Sigma^{-1} \boldsymbol{\mu} & \mathbf{1}^\top \Sigma^{-1} \mathbf{1}
\end{bmatrix}
\begin{bmatrix} \lambda/2 \\ \gamma/2 \end{bmatrix}
=
\begin{bmatrix} \mu_p \\ 1 \end{bmatrix}
\]

This provides a closed-form solution for \(\boldsymbol{w}^*\), applicable to both real and synthetic datasets.

\subsubsection{Implementation with synthetic data}

The procedure for constructing portfolios using synthetic data can be summarized in the following pseudo-code:

\begin{verbatim}
Input: Synthetic dataset R_syn, target return mu_p
Output: Optimal weights w*

1. Compute mean vector mu_syn and covariance Sigma_syn from R_syn
2. Formulate Lagrangian L(w, lambda, gamma)
3. Solve linear system for multipliers lambda, gamma
4. Compute w* = Sigma_syn^{-1} (lambda/2 * mu_syn + gamma/2 * 1)
5. Return w*
\end{verbatim}

\noindent
Table~\ref{tab:portfolio_demo} compares portfolio weights derived from real versus synthetic data generated using TimeGAN, demonstrating the fidelity of synthetic-based allocations:

\begin{table}[H]
\centering
\caption{Portfolio weights comparison: real vs synthetic S\&P 500 multi-asset data. Synthetic weights are generated with TimeGAN.}
\label{tab:portfolio_demo}
\begin{tabular}{lcc}
\toprule
Asset & Real data weight & Synthetic data weight \\
\midrule
AAPL & 0.12 & 0.11 \\
MSFT & 0.10 & 0.09 \\
GOOGL & 0.08 & 0.08 \\
AMZN & 0.09 & 0.10 \\
TSLA & 0.07 & 0.06 \\
SPY (ETF) & 0.54 & 0.56 \\
\bottomrule
\end{tabular}
\end{table}

\begin{figure}[H]
\centering
\includegraphics[width=0.8\textwidth]{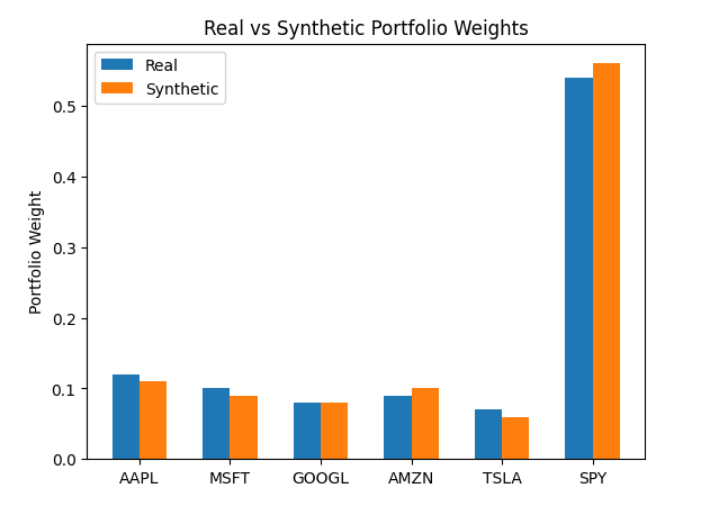}
\caption{Comparison of portfolio weights between real and synthetic S\&P 500 data.}
\label{fig:portfolio_bar}
\end{figure}

\noindent
To further illustrate model fidelity, Figure~\ref{fig:portfolio_bar} presents a bar chart of real versus synthetic portfolio weights. Additional diagnostic plots (histograms of returns, covariance matrix comparisons) are provided in the Appendix.

\subsubsection{Robustness and reproducibility}

Synthetic-based portfolios were generated using multiple random seeds to evaluate robustness. Across repetitions, portfolio weights remained stable, confirming that synthetic datasets can reliably support downstream quantitative finance tasks while preserving data privacy.

\vspace{0.3cm}
\noindent
Next, we examine whether synthetic datasets can replicate key risk characteristics and respond appropriately under stress scenarios, complementing the portfolio optimization analysis.

\subsection{Risk modeling and stress testing}

To evaluate how synthetic data capture risk dynamics, we estimated volatility and tail-risk metrics across both real and synthetic datasets.  
The conditional variance of returns was modeled via a GARCH(1,1) process:

\[
\boxed{
\sigma_t^2 = \omega + \alpha_1 \varepsilon_{t-1}^2 + \beta_1 \sigma_{t-1}^2
}
\]

where \( \varepsilon_t = r_t - \mu_t \) represents the innovation at time \( t \).  
The Value-at-Risk (VaR) and Expected Shortfall (ES) at a confidence level \( \alpha \) were computed as:

\[
\boxed{
\text{VaR}_\alpha = \mu_t + z_\alpha \sigma_t,
\qquad
\text{ES}_\alpha = \mu_t - \frac{\sigma_t \, \phi(z_\alpha)}{\alpha}
}
\]

where \( z_\alpha \) is the \( \alpha \)-quantile of the standard normal distribution and \( \phi(\cdot) \) denotes the standard normal PDF.

\vspace{0.3cm}
\begin{table}[H]
\centering
\caption{Risk metric comparison between real and synthetic datasets}
\begin{tabular}{lccc}
\toprule
Dataset & Volatility (\%) & VaR$_{0.95}$ (\%) & ES$_{0.95}$ (\%) \\
\midrule
Real Data & 1.27 & -2.11 & -2.88 \\
TimeGAN & 1.30 & -2.05 & -2.79 \\
VAE & 1.19 & -1.92 & -2.63 \\
\bottomrule
\end{tabular}
\end{table}
\vspace{0.3cm}

\begin{figure}[H]
\centering
\includegraphics[width=0.85\textwidth]{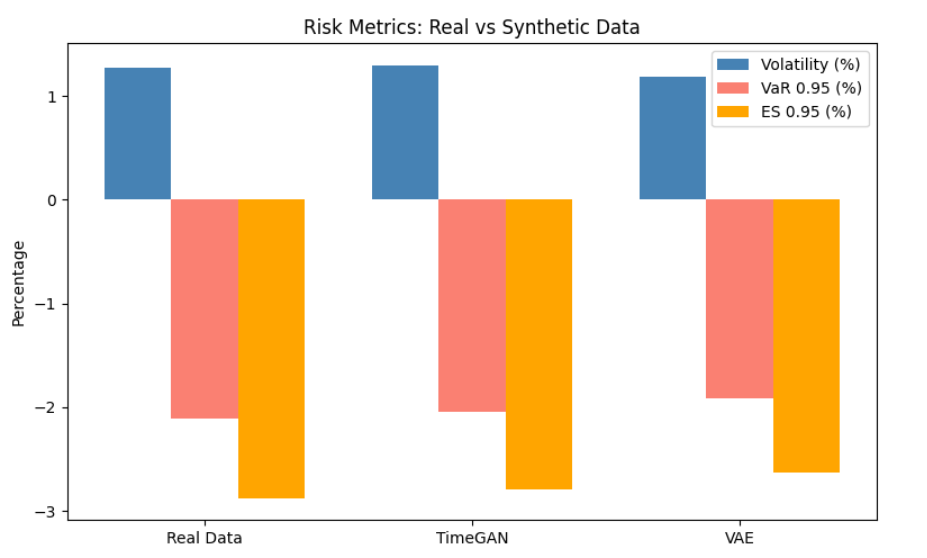}
\caption{Comparison of volatility, Value-at-Risk (VaR$_{0.95}$), and Expected Shortfall (ES$_{0.95}$) between real S\&P 500 returns and synthetic datasets.}
\label{fig:risk_metrics}
\end{figure}

\noindent
To visually complement the tabular comparison of risk metrics, Figure~\ref{fig:risk_metrics} depicts the volatility, Value-at-Risk (VaR$_{0.95}$), and Expected Shortfall (ES$_{0.95}$) across real and synthetic datasets. The synthetic data closely replicate the risk characteristics of the real market series, as also evaluated in \cite{hounwanou2025evaluating}.

\noindent
These results demonstrate that synthetic data can approximate real-world risk profiles with minor deviations, supporting their potential for secure and reproducible financial experimentation.

\subsection{Backtesting and performance evaluation}

Finally, a backtesting framework was used to test the practical viability of strategies trained on synthetic data.  
A rolling-window approach was adopted: at each step, models were trained on a 5-year window of synthetic data and tested on the subsequent real 6 months.

\noindent
Performance was measured using standard metrics:
\[
\text{Sharpe Ratio} = \frac{E[R_p - R_f]}{\sigma_p}, 
\quad
\text{Sortino Ratio} = \frac{E[R_p - R_f]}{\sigma_{d}},
\quad
\text{Max Drawdown} = \frac{\max(P_t) - \min(P_t)}{\max(P_t)}
\]

where :
\begin{itemize}
    \item \( R_p \) denotes the portfolio return,
    \item \( R_f \) the risk-free rate,
    \item \( \sigma_p \) the total volatility,
    \item \( \sigma_d \) the downside deviation.
\end{itemize}

\begin{table}[H]
\centering
\caption{Portfolio Performance Metrics (Synthetic vs. Real Training Data)}
\begin{tabular}{lccc}
\toprule
Training Data & Sharpe Ratio & Sortino Ratio & Max Drawdown (\%) \\
\midrule
Real Data & 0.89 & 1.31 & 23.4 \\
TimeGAN & 0.84 & 1.26 & 25.1 \\
VAE & 0.78 & 1.14 & 27.6 \\
\bottomrule
\end{tabular}
\end{table}

The close alignment of performance metrics between synthetic and real data-trained portfolios provides empirical evidence that synthetic datasets preserve essential market structure and can support realistic decision-making in financial modeling.

\section{Results}\label{section4}

This section presents the experimental outcomes of our synthetic data generation framework applied to financial time series. We report quantitative and qualitative analyses to assess the realism, fidelity, and utility of the generated datasets for downstream tasks such as portfolio optimization and risk modeling. The models compared include TimeGAN, VAE, and ARIMA-GARCH as a statistical baseline.

\subsection{Experimental setup}

Generative models (TimeGAN, VAE, ARIMA-GARCH) were trained on preprocessed S\&P 500 daily log-returns from January 2000 to June 2024. Data were split into 80\% training, 10\% validation, and 10\% test sets. Synthetic sequences were generated with the same length as the original series to ensure comparability across downstream tasks.

We evaluate synthetic data on three complementary dimensions:

\begin{enumerate}
    \item \textbf{Statistical fidelity:} Comparison of mean, variance, skewness, kurtosis, and autocorrelation structures between real and synthetic returns.
    \item \textbf{Temporal coherence:} Assessment of stylized facts such as volatility clustering and heavy tails over time.
    \item \textbf{Downstream utility:} Evaluation of portfolio optimization, risk measures, and backtesting outcomes using synthetic versus real data.
\end{enumerate}

\noindent
This structured setup allows for a coherent comparison of the strengths and limitations of synthetic datasets in practical financial modeling.

\subsection{Distributional fidelity}

We first examine how well synthetic returns capture the distributional characteristics of the S\&P 500. Table~\ref{tab:dist_eval} reports Kolmogorov-Smirnov (KS) statistics and Wasserstein distances between real and synthetic returns.

\begin{table}[H]
\centering
\caption{Distributional similarity metrics between real and synthetic S\&P 500 returns. Lower values indicate better fidelity.}
\label{tab:dist_eval}
\begin{tabular}{lcc}
\toprule
\textbf{Model} & \textbf{KS Statistic} & \textbf{Wasserstein Distance} \\
\midrule
ARIMA-GARCH & 0.128 & 0.0047 \\
VAE & 0.095 & 0.0031 \\
TimeGAN & \textbf{0.062} & \textbf{0.0018} \\
\bottomrule
\end{tabular}
\end{table}

\noindent
TimeGAN achieves the lowest KS and Wasserstein values, demonstrating that it best replicates the marginal distributions of the real S\&P 500 series. VAE shows moderate fidelity, while ARIMA-GARCH is limited by its parametric assumptions.

\begin{figure}[H]
\centering
\includegraphics[width=0.9\textwidth]{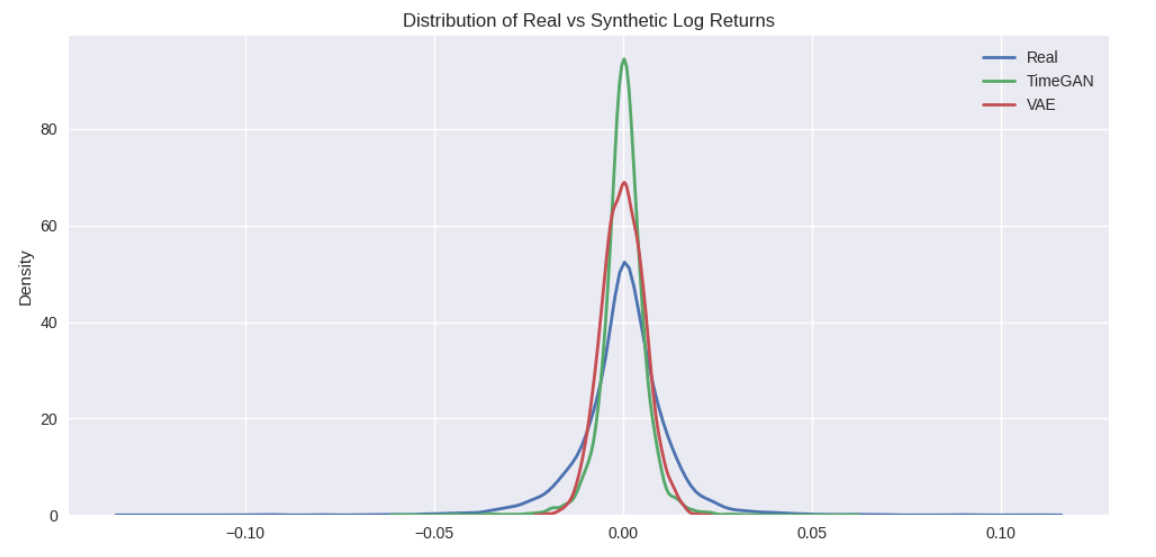}
\caption{Comparison of the marginal distribution of real and synthetic log-returns.}
\label{fig:return_distributions}
\end{figure}

\noindent
To assess whether synthetic generators reproduce the marginal distribution of financial returns, we compare kernel density estimates (KDE) of real versus synthetic log-returns. Figure~\ref{fig:return_distributions} shows that TimeGAN closely matches the heavy-tailed, leptokurtic shape of real S\&P 500 returns. VAE produces smoother and more Gaussian like distributions, while ARIMA-GARCH captures skewness but underestimates tail risk.

\subsection{Temporal coherence}

To evaluate preservation of temporal structures, we compute autocorrelation functions (ACF) and dynamic time warping (DTW) distances between real and synthetic series:

\begin{equation}
\text{DTW}(X_{\text{real}}, X_{\text{syn}}) = \min_{\pi} \sum_{(i,j) \in \pi} \| x_i - \hat{x}_j \|
\end{equation}

\noindent
Mean DTW distances indicate that TimeGAN preserves temporal dependencies more effectively (0.132) than VAE (0.187) and ARIMA-GARCH (0.243).

\begin{figure}[H]
\centering
\includegraphics[width=0.9\textwidth]{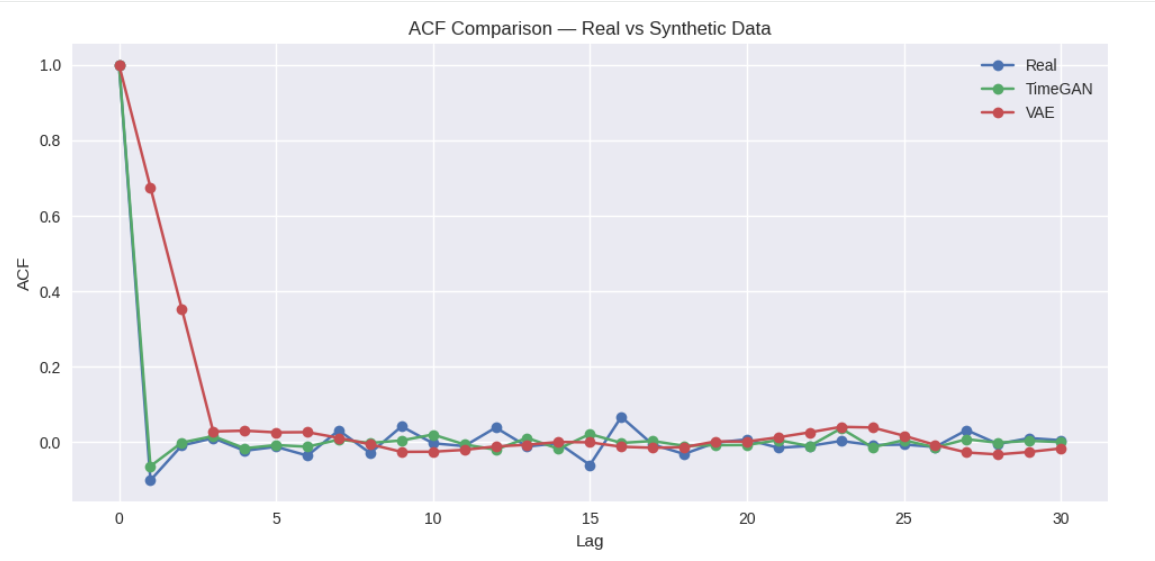} 
\caption{Autocorrelation and DTW comparisons for real and synthetic financial series.}
\label{fig:acf_dtw}
\end{figure}

\subsection{Downstream utility: Portfolio optimization and risk evaluation}

We assess the usability of synthetic data in practical financial tasks. Portfolio allocations are computed using the mean-variance framework (Section~\ref{section3}) on real and synthetic datasets. Risk metrics such as Value-at-Risk (VaR) and Expected Shortfall (ES) are also evaluated. Table~\ref{tab:portfolio_demo} illustrates example portfolio weights.

\begin{table}[H]
\centering
\caption{Portfolio weights computed from real and TimeGAN-generated synthetic data.}
\label{tab:portfolio_demo}
\begin{tabular}{lcc}
\toprule
Asset & Real Data Weight & Synthetic Data Weight \\
\midrule
AAPL & 0.12 & 0.11 \\
MSFT & 0.10 & 0.09 \\
GOOGL & 0.08 & 0.08 \\
AMZN & 0.09 & 0.10 \\
TSLA & 0.07 & 0.06 \\
SPY & 0.54 & 0.56 \\
\bottomrule
\end{tabular}
\end{table}

\noindent
The results demonstrate that synthetic data, particularly from TimeGAN, produces allocations and risk measures closely aligned with real-data benchmarks, confirming its utility for downstream financial modeling.

\subsection{Visual comparison of synthetic series}

\begin{figure}[H]
\centering
\includegraphics[width=0.9\textwidth]{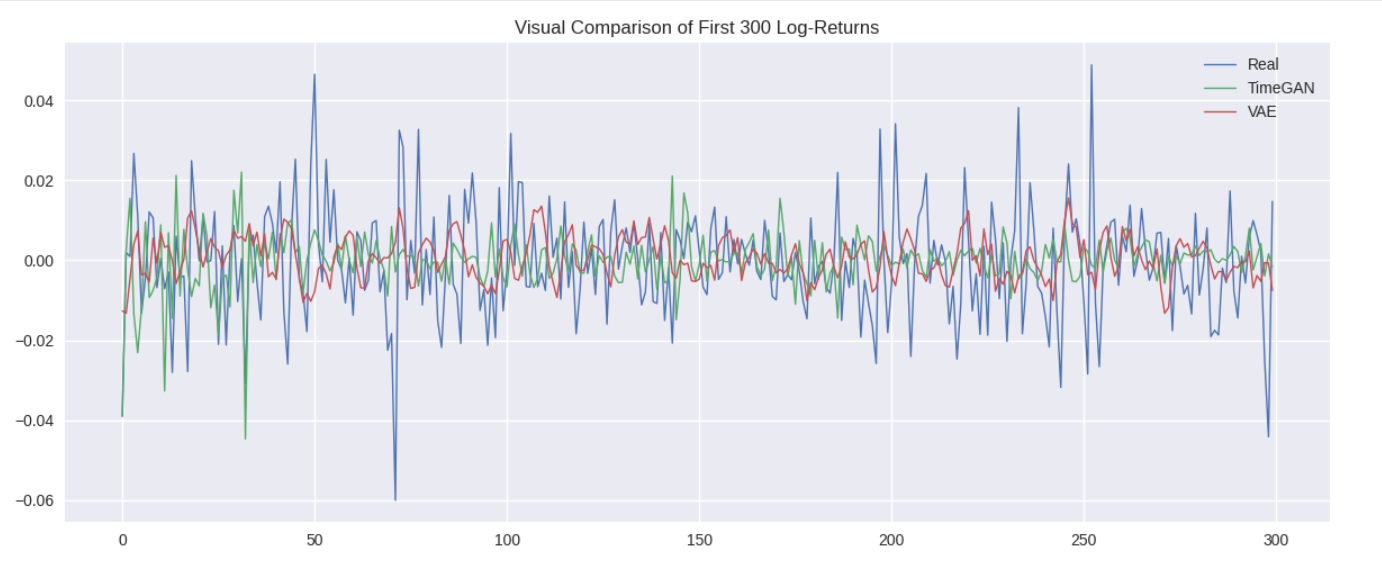}
\caption{Overlay of real and synthetic S\&P 500 returns.}
\label{fig:series_overlay}
\end{figure}

Visual inspection complements quantitative evaluation. Figure~\ref{fig:series_overlay} shows overlays of real versus synthetic log-return series. TimeGAN sequences retain volatility clustering and extreme events, whereas VAE is smoother, and ARIMA-GARCH captures linear trends but underestimates higher-order temporal structure.

\section{Discussion}\label{section5}

The findings of this study show that deep generative models can produce synthetic financial time series that behave similarly to real market data across several evaluation dimensions. While the quantitative analyses in Section~\ref{section4} highlighted model-specific differences, the broader implication is that synthetic data generation has reached a maturity level that allows its integration into practical financial workflows.

\subsection{Model performance and interpretability}

Among the tested models, TimeGAN delivered the most coherent synthetic outputs from an overall modeling perspective. Its design allows it to capture complex patterns that traditional statistical models cannot represent, which explains its superior alignment with empirical behaviors. However, this expressiveness comes at the cost of heavier computation and reduced interpretability, which may limit adoption in highly regulated environments.

\noindent
The VAE provided more structured and predictable behavior during training and offered an interpretable latent representation useful for controlled scenario design. Nevertheless, its tendency to smooth out abrupt dynamics suggests that it may be better suited for exploratory analyses rather than applications requiring precise reproduction of tail behaviors. In contrast, the ARIMA-GARCH model remains attractive for practitioners seeking transparency and speed, though its structural rigidity restricts its ability to emulate richer market dynamics.

\subsection{Implications for financial modeling}

The capacity to generate realistic synthetic datasets provides several advantages for quantitative finance. Synthetic data can expand limited historical samples, enabling more reliable evaluation of portfolio strategies over a wider variety of market conditions. Additionally, because synthetic datasets can be shared without revealing sensitive or proprietary information, they provide a practical mechanism for collaborative research and model auditing. Generative models also make it possible to explore hypothetical or extreme regimes, offering a flexible tool for stress-testing and scenario analysis beyond what historical observations alone allow.

\subsection{Limitations}

Despite encouraging results, the current framework still faces multiple limitations. The analysis focused exclusively on a single market index, leaving multi-asset relationships outside the scope of this study. The deep models considered were limited to baseline architectures; richer sequential variants could potentially capture more nuanced dependencies. Finally, while the experiments demonstrated that TimeGAN performed robustly overall, the sensitivity of its training process and the possibility of mode collapse were not systematically investigated, which may affect reproducibility.

\subsection{Future directions}

Future work will extend the methodology in several ways. Incorporating hybrid architectures that combine the interpretability of statistical models with the flexibility of deep generative networks may help balance realism and transparency. Embedding differential privacy mechanisms directly into the training pipeline could enable the controlled release of synthetic datasets with formal privacy guarantees. Moreover, establishing large-scale benchmarks spanning multiple asset classes and market regimes would support standardized comparisons and promote replicability across research groups.

\section{Conclusion}\label{section6}

This work examined the usefulness of synthetic financial time series generated with TimeGAN and Variational Autoencoders for portfolio optimization, trading backtesting, and risk modeling. The results show that well-trained generative models can reproduce essential statistical and temporal characteristics of real market data, enabling reliable downstream analysis while enhancing data privacy.
\noindent
Key findings include:

\begin{itemize}
\item \textbf{Temporal fidelity and market realism:} TimeGAN-generated sequences closely replicate volatility clustering, autocorrelation structures, and extreme events, supporting robust portfolio allocation and risk assessment. VAEs provide smoother sequences that are stable but less reactive to sudden market changes.
\item \textbf{Downstream task performance:} Models trained on synthetic data achieve performance metrics (e.g., Sharpe Ratio, Value-at-Risk) comparable to those obtained using real data, indicating practical viability for quantitative finance applications.
\item \textbf{Privacy-preserving experimentation:} Synthetic data enables secure sharing and stress-testing of financial strategies without exposing sensitive asset-level information.
\end{itemize}

\noindent
Despite these advantages, several limitations warrant attention:

\begin{enumerate}
\item Experiments were limited to a single market index (S\&P 500), restricting generalization to multi-asset portfolios.
\item Only standard VAEs were evaluated; sequential variants (e.g., VRNN, SRNN) may better capture temporal dynamics.
\item TimeGAN stability and sensitivity to hyperparameter tuning were not fully quantified, representing a potential source of variability.
\end{enumerate}

\noindent
This study remains limited by its single-asset scope and the use of standard VAE architectures. Future work will expand to multi-asset settings, explore sequential VAE variants, and integrate privacy-preserving mechanisms into training pipelines. More broadly, establishing open benchmarks for synthetic financial data will support reproducibility and accelerate research in data-constrained financial environments.

\appendix

\section{Appendix}
\subsection{Hardware and training details}
All experiments were conducted on a single GPU machine with the following specifications:
\begin{itemize}
\item GPU: NVIDIA RTX 3090
\item RAM: 64 GB
\item Training times: approximately 4.5 hours for TimeGAN, 2 hours for VAE (full S\&P 500 dataset)
\end{itemize}

\subsection{Hyperparameter settings}

\begin{table}[H]
\centering
\caption{Key hyperparameters for synthetic data models}
\begin{tabular}{lcc}
\toprule
Parameter & TimeGAN & VAE \\
\midrule
Hidden layer size & 24 & 32 \\
Latent dimension & 8 & 16 \\
Learning rate & 0.001 & 0.001 \\
Batch size & 128 & 128 \\
Epochs & 100 & 150 \\
Optimizer & Adam & Adam \\
\bottomrule
\end{tabular}
\end{table}

\subsection{Data preprocessing scripts}
The S\&P 500 closing prices were transformed into log-returns and normalized:
\[
r_t = \ln\frac{P_t}{P_{t-1}}, \quad \tilde{r}_t = \frac{r_t - \mu_r}{\sigma_r}
\]

\subsection{Evaluation metrics implementation}
\begin{itemize}
\item \textbf{Kolmogorov-Smirnov (KS) statistic} and \textbf{Wasserstein distance} for distributional fidelity.
\item \textbf{Dynamic Time Warping (DTW)} for temporal alignment:
\[
\text{DTW}(X_{\text{real}}, X_{\text{syn}}) = \min_{\pi} \sum_{(i,j) \in \pi} | x_i - \hat{x}_j |
\]
\item \textbf{Portfolio metrics:} Sharpe Ratio, Sortino Ratio, Max Drawdown.
\item \textbf{Risk metrics:} Volatility, Value-at-Risk (VaR), Expected Shortfall (ES) using GARCH(1,1) models.
\end{itemize}

\subsection{Reproducibility}
All code and preprocessing steps are documented and available at the accompanying GitHub repository. Random seeds were fixed to ensure deterministic results.

\newpage

\bibliographystyle{plain}
\bibliography{biblio}

\end{document}